\documentclass[aps,prl,twocolumn]{revtex4-2}
\pdfoutput=1

\usepackage{amsmath}
\usepackage{amssymb}
\usepackage{amsbsy}
\usepackage{mathtools}
\usepackage{dsfont}
\usepackage{hyperref}
\usepackage{xcolor}

\definecolor{darkred}{RGB}{139,0,0}
\definecolor{darkgreen}{RGB}{0,100,0}
\definecolor{darkblue}{RGB}{6,69,173}

\hypersetup{
breaklinks=true,
colorlinks,
linkcolor=darkblue,
filecolor=darkgreen,
urlcolor=darkblue,
citecolor=darkblue}

\newcommand{\ev}[1]{\ensuremath{\langle #1\rangle}}
\newcommand{\ket}[1]{\ensuremath{|#1\rangle}}
\newcommand{\kobra}[1]{\ensuremath{|#1\rangle\langle #1|}}
\newcommand{\nn}{\nonumber}
\newcommand{\tr}{\operatorname{Tr}}
\newcommand{\re}{\operatorname{Re}}
\newcommand{\cov}{\operatorname{Cov}}

\begin{document}

\noindent\textbf{Comment on ``Fluctuations in Extractable Work Bound the Charging Power of Quantum Batteries''}

In Ref.~\cite{Garcia-Pintos:2020fi} the connection between the charging power of a quantum battery and the fluctuations of a ``free energy operator'' whose expectation value characterizes the maximum extractable work of the battery is studied using both closed- and open-system analyses. 
Recently, it has been shown~\cite{Cusumano:2021co} that the conclusions of Ref.~\cite{Garcia-Pintos:2020fi} do \emph{not} hold for open-system dynamics. 
Since the two analyses are physically equivalent approaches to studying the dynamics of the battery, in light of the findings of Ref.~\cite{Cusumano:2021co} we critically examine whether the conclusions of Ref.~\cite{Garcia-Pintos:2020fi} hold for closed-system dynamics.
In doing so, we find a few mistakes in the analysis of Ref.~\cite{Garcia-Pintos:2020fi} and obtain the correct bound on the charging power. 
As a result, for closed-system dynamics the conclusions of Ref.~\cite{Garcia-Pintos:2020fi} are in general not correct.

The starting point of our analysis is the charging power $P(t)$ of the battery, given by Eq.~(8) of Ref.~\cite{Garcia-Pintos:2020fi} as
\begin{equation}
P(t)=-i\tr([\rho,\mathcal{F}\otimes\openone_\mathcal{SBA}]V),
\end{equation}
where $\rho$ is the full state of the closed system $\mathcal{SBAW}$, $\mathcal{F}$ is a Hermitian operator, known as the free energy operator of the battery $\mathcal{W}$, and $V$ is the interaction Hamiltonian between the battery and the bath $\mathcal{B}$, system $\mathcal{S}$, and ancilla $\mathcal{A}$. 
Following Ref.~\cite{Garcia-Pintos:2020fi}, we define $\delta\mathcal{F}=\mathcal{F}-\ev{\mathcal{F}}_\mathcal{W}$ and $\delta V=V-\ev{V}$, where $\ev{\mathcal{F}}_\mathcal{W}=\tr(\rho_\mathcal{W}\mathcal{F})$ and $\ev{V}=\tr(\rho V)$. After some algebra, we obtain
\begin{equation}
|P(t)|^2=|\tr(\rho[\delta\mathcal{F},\delta V])|^2,\label{eq:Poftsq1}
\end{equation}
where for notational simplicity we will henceforth use the shorthand notation $\delta\mathcal{F}=\delta\mathcal{F}\otimes\openone_\mathcal{SBA}$.
Note that Eq.~\eqref{eq:Poftsq1} is an equality instead of an inequality given by Eq.~(9) of Ref.~\cite{Garcia-Pintos:2020fi}. 
It is convenient to rewrite Eq.~\eqref{eq:Poftsq1} as
\begin{equation}
|P(t)|^2=|\tr(\sqrt{\rho}\,\delta\mathcal{F}\delta V\sqrt{\rho}-\sqrt{\rho}\,\delta V\delta\mathcal{F}\sqrt{\rho})|^2,\label{eq:Poftsq2}
\end{equation}
where we have used the fact that $\rho$ is a positive operator.
We note that since $\sqrt{\rho}\,\delta\mathcal{F}\delta V\sqrt{\rho}$ and $\sqrt{\rho}\,\delta V\delta\mathcal{F}\sqrt{\rho}$ are Hermitian conjugates of each other, it follows that $\tr(\sqrt{\rho}\,\delta\mathcal{F}\delta V\sqrt{\rho})$ and $\tr(\sqrt{\rho}\,\delta V\delta\mathcal{F}\sqrt{\rho})$ are complex conjugates of each other. 
As a matter of fact, this is the utmost important point that is missed in the analysis of Ref.~\cite{Garcia-Pintos:2020fi}. With this point in mind, we can rewrite Eq.~\eqref{eq:Poftsq2} as
\begin{eqnarray}
|P(t)|^2&=&|\tr(\sqrt{\rho}\,\delta\mathcal{F}\delta V\sqrt{\rho})|^2+|\tr(\sqrt{\rho}\,\delta V\delta\mathcal{F}\sqrt{\rho})|^2\nn\\
&&-2\re([\tr(\rho\,\delta\mathcal{F}\delta V)]^2),\label{eq:Poftsq3}
\end{eqnarray}
where Re denotes the real part.

To find the upper bound on $|P(t)|^2$, again following Ref.~\cite{Garcia-Pintos:2020fi}, we use the fact that for a positive operator $A$ and Hermitian operators $B$ and $C$, the Cauchy-Schwarz inequality implies $|\tr(\sqrt{A}BC\sqrt{A})|^2\le|\tr(AB^2)|\,|\tr(AC^2)|$.
Equation~\eqref{eq:Poftsq3} then leads to  
\begin{eqnarray}
|P(t)|^2&\le&2\bigl([\tr[\rho(\delta\mathcal{F})^2]\tr[\rho(\delta V)^2]-\re([\tr(\rho\,\delta\mathcal{F}\delta V)]^2)\bigl)\nn\\
&=&2\bigl(\sigma^2_\mathcal{F}\sigma^2_V-\re[\cov(\mathcal{F},V)^2]\bigr).\label{eq:Poftsqineq}
\end{eqnarray}
Here, $\sigma^2_\mathcal{F}$ is the variance of $\mathcal{F}$ in the battery state $\rho_\mathcal{W}$, $\sigma^2_V$ is the variance of $V$ in the full state $\rho$, and $\cov(\mathcal{F},V)$ is the covariance between $\mathcal{F}$ and $V$ in the full state $\rho$. Specifically, we have
\begin{equation}
\begin{gathered}
\sigma^2_\mathcal{F}=\ev{\mathcal{F}^2}_\mathcal{W}-\ev{\mathcal{F}}^2_\mathcal{W},\quad
\sigma^2_V=\ev{V^2}-\ev{V}^2,\\
\cov(\mathcal{F},V)=\ev{(\mathcal{F}\otimes\openone_\mathcal{SBA})V}-\ev{\mathcal{F}}_\mathcal{W}\ev{V}.
\end{gathered}
\end{equation}
Moreover, the Cauchy-Schwarz inequality $\sigma^2_\mathcal{F}\sigma^2_V\ge|\cov(\mathcal{F},V)|^2$ implies $\sigma^2_\mathcal{F}\sigma^2_V-\re[\cov(\mathcal{F},V)^2]\ge 0$ as it should be.
Evidently, Eq.~\eqref{eq:Poftsqineq} is the corrected expression for Eqs.~(9) and (12) of Ref.~\cite{Garcia-Pintos:2020fi}.

The last step of our analysis is to verify the case in which the battery state is an eigenstate of the free energy operator.  
Suppose $\rho_\mathcal{W}=\kobra{j}$ and $\mathcal{F}\ket{j}=w_j\ket{j}$ with $w_j$ the real eigenvalue;
we obtain $\sigma^2_\mathcal{F}=\cov(\mathcal{F},V)=0$, which implies $P(t)=0$.
However, under the assumption of a very general charging process with $\sigma^2_V\ne 0$, we stress that in our analysis $\rho_\mathcal{W}=\kobra{j}$ is only a \emph{sufficient} condition for the battery to have a nonzero charging power, as opposed to a \emph{sufficient and necessary} condition in the original incorrect analysis of Ref.~\cite{Garcia-Pintos:2020fi}. 
Moreover, even though the total system is initially in a product state with the battery in an eigenstate of $\mathcal{F}$, the interaction $V$ will make the battery entangled with the other subsystems, giving rise to a mixed battery state.   
It is conceivable that there exist entangled full states $\rho$ and mixed battery states $\rho_\mathcal{W}=\tr_\mathcal{SBA}(\rho)$ with nonzero $\sigma^2_\mathcal{F}$ and $\cov(\mathcal{F},V)$ but $P(t)=0$. 

Since there is still no consensus on the notion of work in the quantum regime, the discrepancy between the conclusions of Ref.~\cite{Cusumano:2021co} and of the present work seems to suggest that the free energy operator $\mathcal{F}$ introduced in Ref.~\cite{Garcia-Pintos:2020fi} does not properly quantify the work content of the battery in a physically consistent manner. 
This is certainly an important issue that deserves further investigation.

This work was supported in part by the MOST of Taiwan under grant 109-2112-M-032-009.

\vspace{3ex}
\noindent Shang-Yung Wang\\
\small{\indent Department of Physics\\
\indent Tamkang University\\
\indent New Taipei City 25137, Taiwan}


\end{document}